# Radio-frequency impedance measurements using a tunnel-diode oscillator (TDO) technique

H. Srikanth*, J. Wiggins and H. Rees
*Advanced Materials Research Institute, University of New Orleans, New Orleans LA 70148*

A resonant method based on a tunnel-diode oscillator (TDO) for precision measurements of relative impedance changes in materials, is described. The system consists of an effective self-resonant LC-tank circuit driven by a forward-biased tunnel diode operating in its negative resistance region. Samples under investigation are placed in the core of an inductive coil and impedance changes are determined directly from the measured shift in resonance frequency. A customized low temperature insert is used to integrate this experiment with a commercial Model 6000 Physical Property Measurement System (Quantum Design). Test measurements on a manganese-based perovskite sample exhibiting colossal magneto-resistance (CMR) indicate that this method is well suited to study the magneto-impedance in these materials.

## 1. INTRODUCTION

Resonant methods have the advantage of precision and high sensitivity when it comes to detecting changes in the physical properties of materials as a function of temperature and magnetic field. This is ascribed to the fact that frequency can be measured with a very high degree of accuracy. In a typical resonant technique based on an LC-tank circuit, the capacitor or inductor couples to the material under study and act as transducers of physical parameters. In other words, any change in material properties will induce a change in the capacitance or inductance, which in turn results in a shift in the resonance frequency. So, measurement of the frequency shift translates to directly probing the electronic, dielectric or magnetic response of the material to the oscillating signal.

Tunnel-diode oscillators (TDO) which essentially operate based on the principle outlined above, have been used in the past to study a variety of issues ranging from paramagnetic susceptibility in salts to penetration depth in superconductors [1-4]. Over the past few years, the Northeastern University group has demonstrated the success of using this method to investigate phenomena in new materials like cuprate superconductors [5], borocarbide superconductors [6] and more recently, manganites exhibiting colossal magneto-resistance (CMR) effects [7].

While there are some limitations associated with the inherent design of TDO-based resonant methods in the elucidation of physical properties of materials, these are overshadowed by several advantages over standard DC methods. It is clear that this is proving to be a unique and excellent way to study collective phenomena associated with spin and charge dynamics in novel electronic materials.

In this paper, we describe a TDO-based impedance measurement system that we have built at the Advanced Materials Research Institute, University of New Orleans. Our design includes several improvements over previous instruments and in particular, for the first time, we demonstrate the integration of this measurement system with the commercially available Physical Property Measurement System (Quantum Design). Design and operation of the circuit, customized low temperature probe, software interface and test measurements are described below.

## 2. CIRCUIT DESIGN AND OPERATION

The principle of the TDO can be briefly described as follows. An LC-tank circuit is maintained at a constant amplitude resonance by supplying the circuit with external power to compensate for dissipation. This power is provided by a tunnel diode that is precisely forward biased with a voltage in the region of negative slope of its I-V characteristic, or the so-called negative resistance region. Such an arrangement makes it a self-resonant circuit as the power supplied by the diode maintains continuous oscillation of the LC-tank operating at a frequency given by the standard expression,

$$w = \frac{1}{\sqrt{LC}} \qquad (1)$$

* Corresponding author; e-mail: sharihar@uno.edu

When a sample is inserted into the oscillator tank coil, there is a small change in the coil inductance ΔL. If ΔL/L << 1, one can differentiate equation (1) and obtain the expression:

$$\frac{\Delta w}{w} \approx -\frac{\Delta L}{2L} \qquad (2)$$

The inductance change is related to material properties.

For example, in the case of a magnetic material [2], this is proportional to the real part μ' of the complex permeability,

$$m = m' - im'' \qquad (3)$$

In addition to the general relationships outlined above, if the frequency of oscillation is in the RF range, we have to also consider classical electrodynamics associated with the skin depth [8]. This is particularly important when one is dealing with metallic samples. In this case, the complex impedance of a metal is given by,

$$Z = (1 + i)\frac{r}{d} \qquad (4)$$

where ρ is the resistivity and δ is the skin depth.
From classical electrodynamics, it can be shown that the change in inductance is directly related to the change in impedance, as follows:

$$\Delta L \propto \Delta Z \propto \Delta\left(\sqrt{mr}\right) \qquad (5)$$

where ρ represents the resistivity and μ the permeability of the material.
It is clear from the expression that a measurement of the resonance frequency shift Δω will reflect changes in both the material resistivity and permeability. Such a measurement of a coupled quantity has distinct advantages and forms the basis of probing the spin and charge dynamics in a single experiment. This feature is attractive when looking at correlated materials like CMR oxides that show a strong tendency for interplay among their structural, electronic and magnetic properties [9]. Indeed, the TDO technique is well suited to study phase transitions in these materials as was recently shown by Srikanth et al. [7].

A schematic of the TDO circuit designed by us is shown in Fig. 1. The LC circuit forming the tank resonator is shown in the dashed box. The semi-rigid co-axial cable segment shown in the diagram represents a 4-foot length of RG402 cable (Microcoax) to one end of which the inductive coil L is attached. This cable is directly attached to a low temperature cryostat that enables the coil alone to be introduced into a cryogenic environment. Samples under study are placed inside the core of the inductive coil. Details on the low temperature probe is presented in the next section. The rest of the circuit and the measurement instruments are all located at room temperature.

The tunnel diode or back diode (Model BD 4, Germanium Power Devices) supplies power to the LC-tank and is forward biased as shown in the diagram using a stable digitally programmable DC power supply (Hewlett-Packard Model 6612C). The resistors $R_1$ and $R_2$ make up a voltage divider and $C_1$ is a bypass capacitor. A surface mount monolithic microwave amplifier (Model MAR-8, Mini-Circuits) with a 20dB gain is added as a pre-amplifier at the output stage as indicated. We find this is essential to properly trigger the frequency counter. The values of $R_1$ and $R_2$ were appropriately chosen so that the same DC power supply can be used to provide the correct bias to the tunnel diode and also power the microwave amplifier. A high-resolution frequency counter (Hewlett-Packard Model 53181A) is used to read the resonant frequency and the output signal is also monitored with a digital oscilloscope (LeCroy).

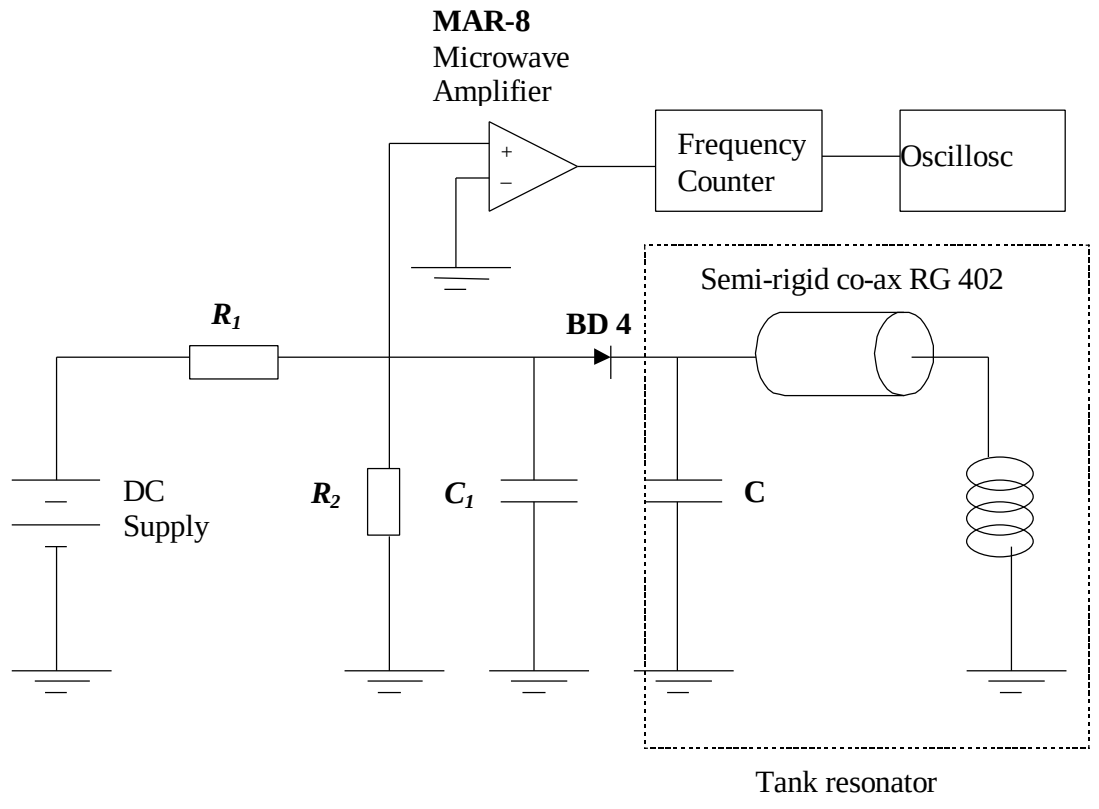

Fig. 1 Schematic of the TDO circuit

The circuit components are chosen with care not only to determine the operating frequency but also to achieve optimum conditions for sustained stable oscillations. There are limitations associated with the resistance of the inductive coil and external capacitance of the co-axial cables used in the system. Choice of circuit component values crucially depends on the ability to compensate for these unavoidable features.

The tank inductance L and capacitance C are vital components of the TDO and have to be chosen with utmost care. For C, surface mount high-Q RF chip capacitors (American Technical Ceramics) with values ranging from 470pF to 1000pF were tried out. Note that it is important to have the value of C higher than the capacitance of the 4-foot RG402 co-axial assembly which is rated at 29pF/foot. The inductive coil is a 20-turn solenoid hand-wound using AWG42 insulated Cu wire around a hollow ceramic tube (0.5cm diameter) and potted with polystyrene epoxy resin. A coating of GE varnish is also applied for good thermal conductivity. Coils with various diameters and wire thickness were tested for stable performance of the tank circuit. Best results were obtained for typical inductance values of L = 1 to 5 μH.

The LC circuit should be considered as an L-R-C circuit because of the inherent internal resistance of the inductive coil. Due to this fact, a damping factor should be taken into account, given by

$$g = \frac{R}{2L} \qquad (6)$$

If the resonance frequency, $\omega >> \gamma$, the output signal has a well-defined sinusoidal form, whereas the signal is distorted from a sinusoidal wave if $\omega$ and $\gamma$ are comparable. A high damping factor with an improper operating current for a tunnel diode oscillator, results in a complex waveform. When non-sinusoidal and highly complex waves are generated, the frequency counter may read erroneous values. Our design took this into consideration and the coil was appropriately made to realize a clean sinusoidal output signal.

There is another important parameter involved in the TDO design consideration. This is the quality factor (Q) defined as:

$$Q = \frac{w\ L}{R} \qquad (7)$$

A high Q is beneficial for stable oscillations. The constraints set by desirable high Q and low damping coupled with the unavoidable capacitance of the long co-axial cable set the upper limit for the resonance frequency in our TDO design.

In Fig. 1, except for the coil (L), the DC power supply and the measurement instruments (frequency counter and oscilloscope), all the other circuit components are mounted on a PCB and enclosed inside an Aluminum box with BNC ports provided for the input and output. A crucial aspect of any RF circuit design is having a proper ground plane. During our circuit assembly and testing,, some of the main problems we encountered had to do with improper shielding. We had to change the layout of the components on the PCB and also make sure there is a uniform ground plane that is also connected to the Al chassis.

Fig. 2 shows the output signal trace from the storage oscilloscope with the TDO operating at a resonance frequency around 6 MHz. A clean nearly sinusoidal waveform is apparent with a peak-to-peak amplitude of around 170mV. The amplitude can be adjusted by tuning the diode bias voltage in the narrow region of its negative resistance characteristic. For stable operation, it is preferable to set the bias closer to the middle of the negative slope in the I-V characteristic. The bias values were taken from the specification sheet provided by the manufacturer of the BD 4 diode (Germanium Power Devices) and we also independently confirmed the tunnel diode operation using a curve tracer.

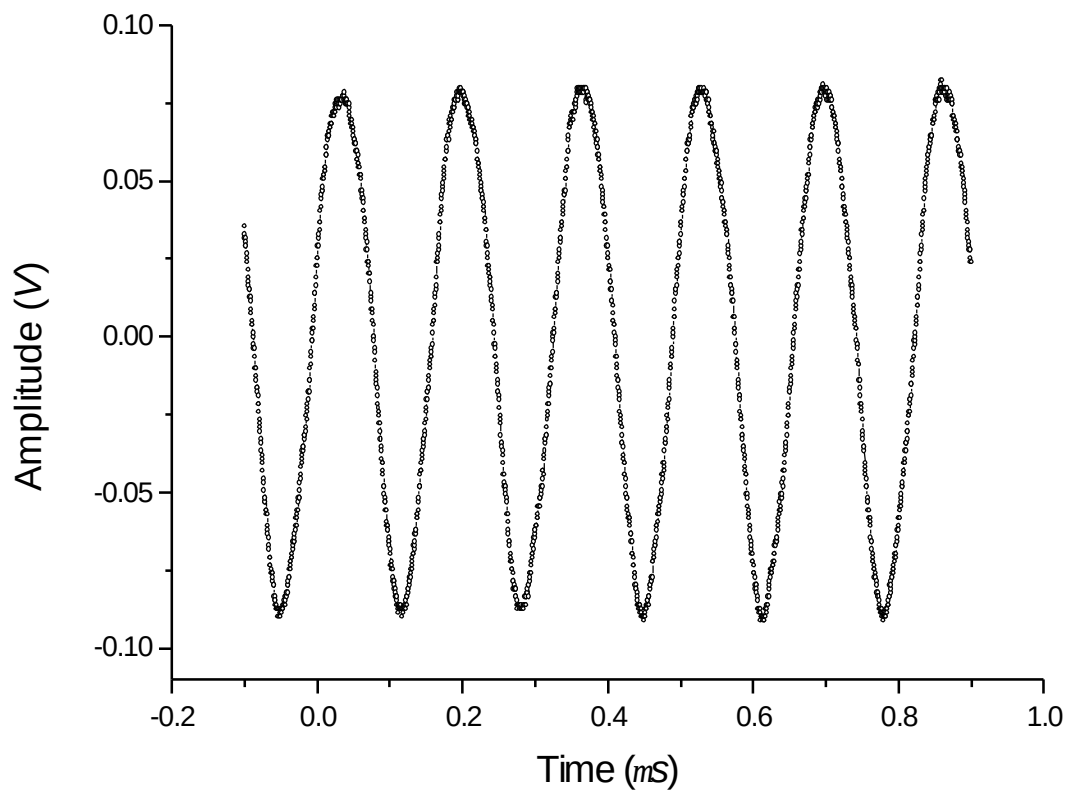

Fig. 2 The output waveform from the TDO circuit measured with a digital oscilloscope. The resonance frequency is around 6 MHz.

Short and long term stability tests of the TDO were conducted and our circuit showed excellent stability. Drift in the resonance frequency (~ 6MHz) is limited to 2 to 3 Hz over a period of 20 to 30 minutes. Tests over a 24-hour period established an overall drift around 500Hz. It is difficult to track down the precise source for these drifts as several factors can contribute. Some possible candidates include drift in the bias supply, diode operation, thermal dissipation inside the confined Al box mainly from the microwave amplifier. Nevertheless, these small drifts are at least a few orders of magnitude lesser than typical resonance frequency shifts encountered in a measurement and do not affect the results.

## 3. LOW TEMPERATURE CRYOSTAT

The versatility of the TDO experiment is greatly enhanced when it is adapted to conduct measurements on materials over a wide range in temperature and magnetic field. This is achieved in our system through integration of our home-built TDO circuitry with a customized cryogenic user probe that fits into the bore of a commercial Physical Property Measurement System (Quantum Design). A semi-rigid co-axial cable (RG402) assembly is attached to the probe. The inductive coil L from the circuit shown in Fig. 1, is connected to the bottom end of the co-axial cable while the top end is terminated by an SMA female connector that can be mated to the rest of the TDO circuit. A schematic of the lower end of the co-axial probe that is in the temperature and field controlled region of the PPMS is shown in Fig. 3.

Samples are placed in gelcaps and inserted into the resonant coil. They are securely fastened with Teflon tape to ensure rigidity. The sample and coil are in thermal contact with the temperature sensor (Cernox) provided on the Quantum Design user probe. Thermal contact is made using Apiezon “N” grease. The co-ax is fixed in the central bore of the user probe. A double “O” ring seal at the top flange (not shown) is used to maintain a vacuum seal around the co-ax cable. The inner conductor and the outer shielding are used for electrical connections. Such an arrangement requires only one cable but the co-ax must be electrically isolated from the surrounding environment. Electrical isolation was obtained by wrapping the co-ax in Teflon tape and placing a small rubber “O” ring close to the base of the co-ax. Once the sample and coax were mounted in the user probe, the probe is then inserted into the PPMS. The PPMS is then used a platform for varying the sample temperature (5K < T < 350K) and static magnetic field (0 < H < 9T).

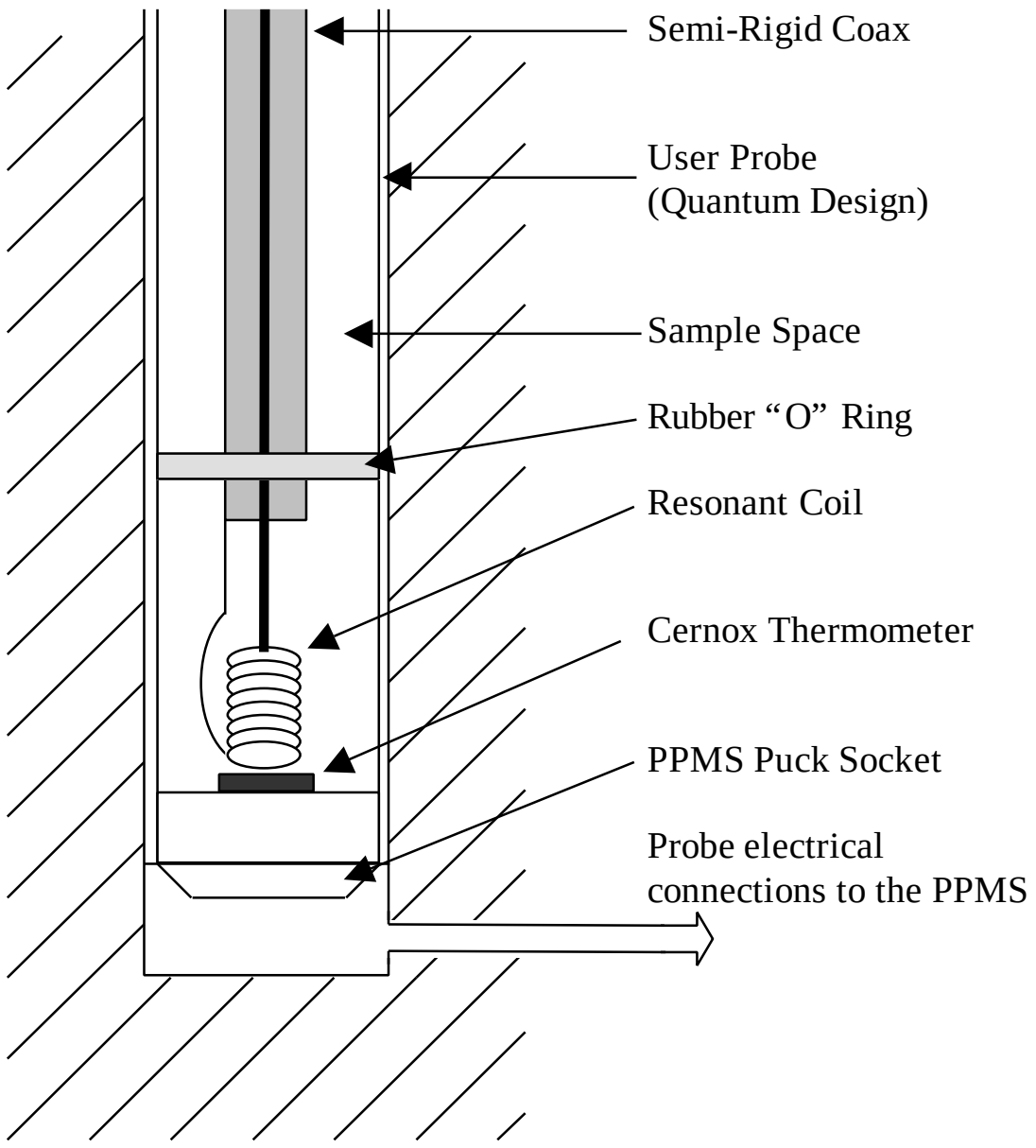

Fig. 3 Schematic of the low temperature probe showing the sample region

Thus, our design combines the ease of operation of the PPMS (including the possibility of changing samples without warming up the system) and its versatile temperature and static field control, with the TDO set-up.

## 4. COMPUTER INTERFACE AND DATA ACQUISITION

Control of the PPMS measurement environment was through using the standard software and computer supplied with the PPMS. Sequences can be written that would control the temperature and field as a function of time. The analog output ports of the PPMS were utilized as monitors for the field and temperature. The signals were then connected to the AUX Input ports of an SR844 lock-in amplifier that served as an A/D converter. It should be noted that this was only used to convert analog levels to GPIB readable data. This data is read via GPIB interface using a second data acquisition computer that also communicates with the frequency counter at the TDO output stage.

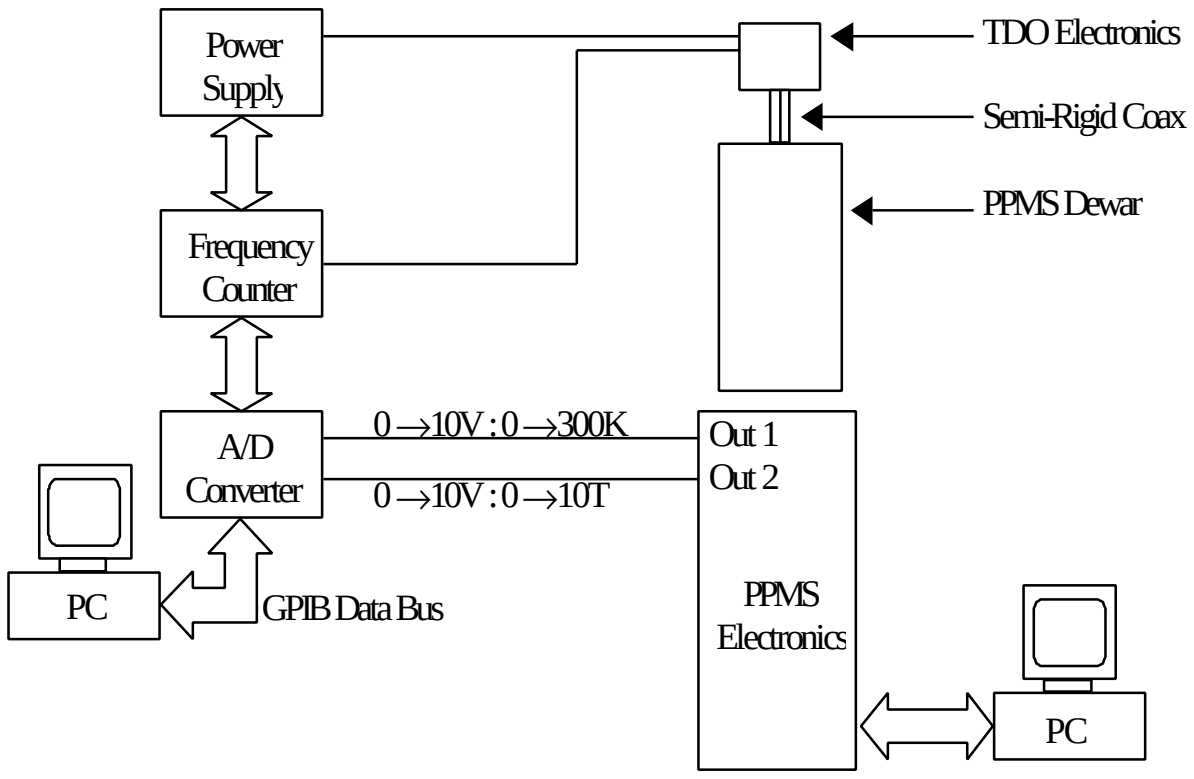

Fig. 4 Schematic of the TDO measurement system

A typical run would have the PPMS scan either magnetic field or temperature over a certain range. The TDO data acquisition computer would then monitor the GPIB data bus reading temperature, field and resonant frequency. A schematic of this arrangement is indicated in Fig. 4.

## 5. TEST MEASUREMENTS

In this section, we present preliminary test measurements using our TDO setup operating in tandem with the PPMS. Prior to measurements on samples, the temperature and field dependence of the empty coil alone (i.e. no sample loaded in its core) is mapped out. This data is shown in Fig. 5. The main panel displays a plot of the resonance frequency shift represented as a dimensionless quantity ($\Delta\omega/\omega$), as the empty coil is cooled in zero field from room temperature to around 20K and the inset shows the field dependence up to 3T at a fixed temperature of 100K. The reasoning behind plotting $\Delta\omega/\omega$ as presented in the figure is as follows. At room temperature, the resonance frequency of the TDO is $\omega_0$ = 5.52226 MHz and as the coil cools down, this frequency monotonically *increases* to 5.52364 MHz at
20K, a change of 1380 Hz. So, in the figure we have effectively represented the frequency shift as $(\omega_0 - \omega)/\omega_0$ and hence the negative values. From equations (1)-(5) in section 2, it can be deduced that this convention of plotting the frequency shift is a meaningful one and helps to visualize the data in terms of impedance changes associated with material properties like resistivity or permeability. Negative or positive signs for the quantity ($\Delta\omega/\omega$) represent decrease or increase in impedance with respect to the reference value of zero.

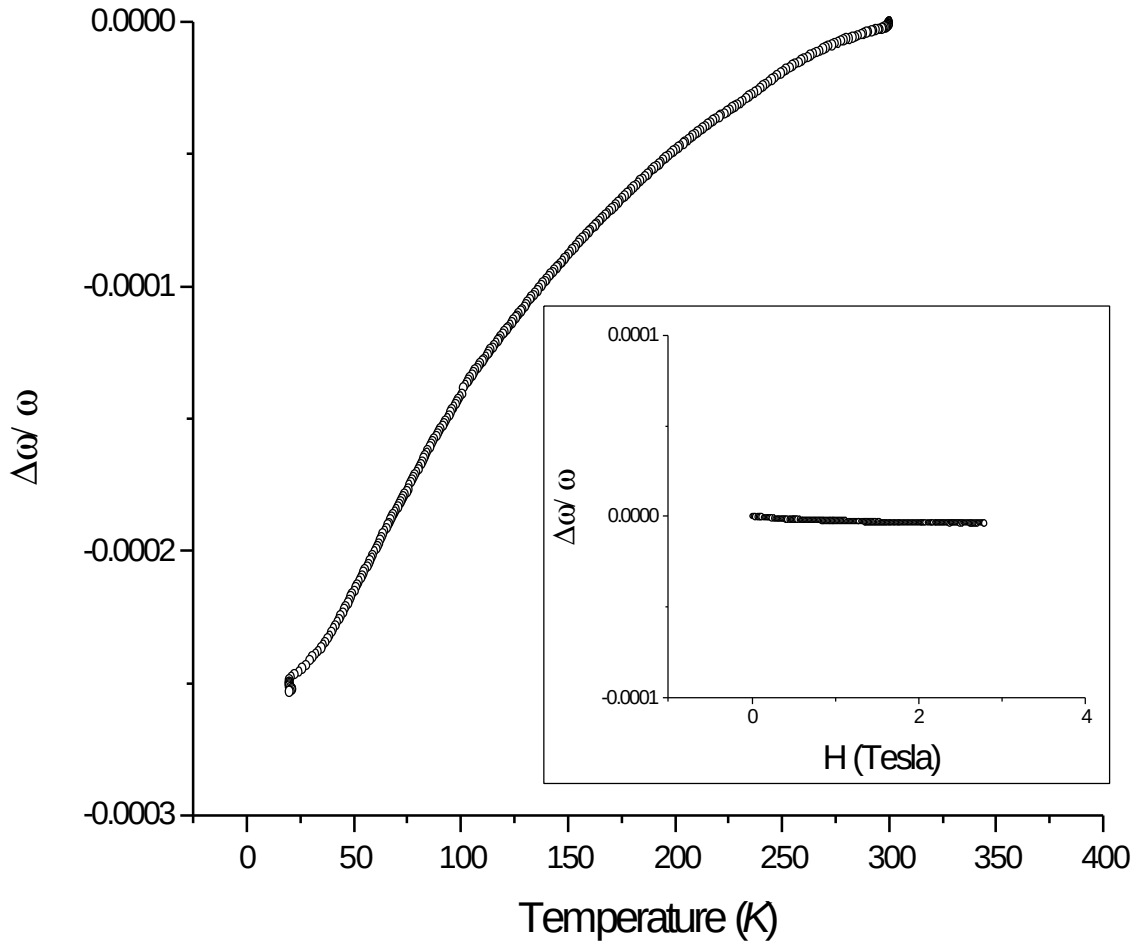

Fig. 5 The RF shift as a function of T for the empty coil. Inset shows the field dependence at a fixed temperature of 100K.

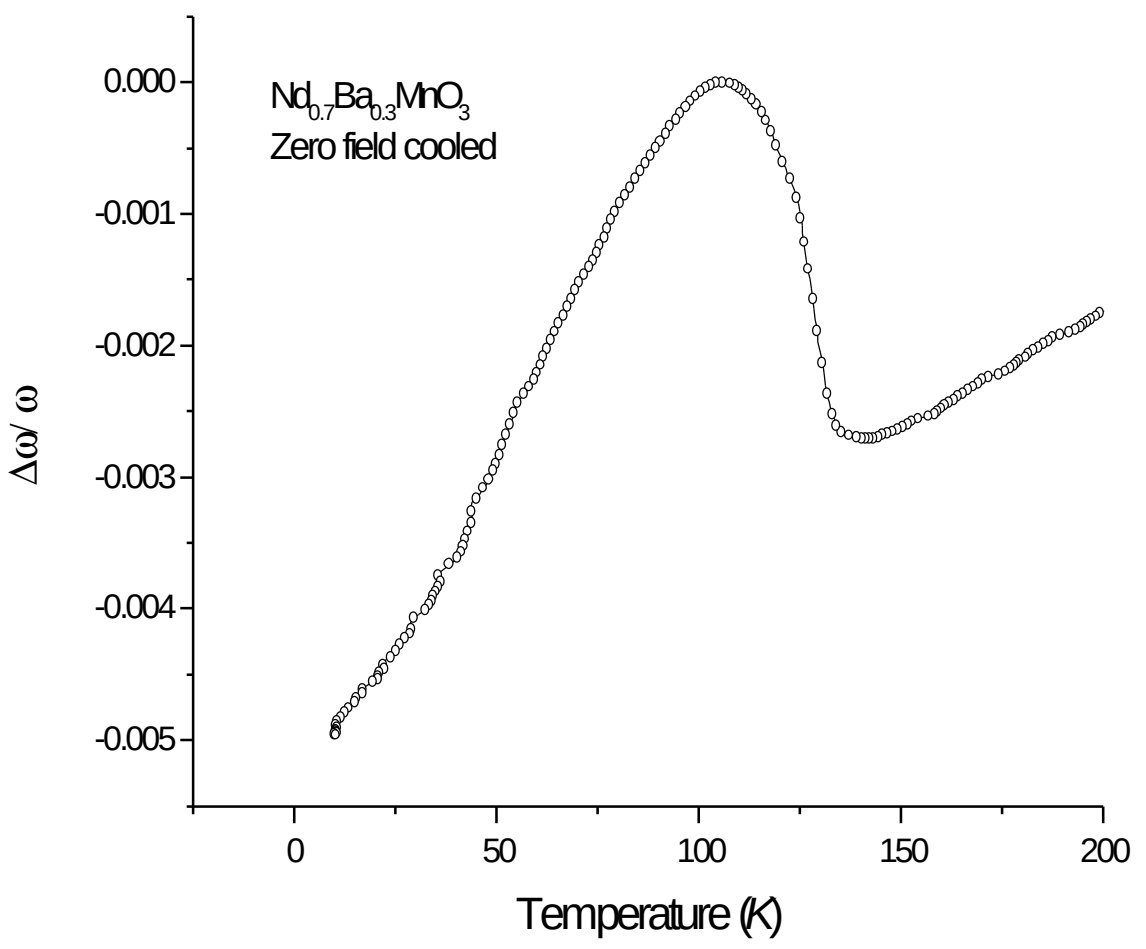

Fig. 6 Measured zero-field temperature dependence of the RF shift for an Nd-Ba-Mn-O sample.

The temperature dependence shown in Fig. 5 is a combined result of decrease in resistivity of the Cu wire making up the coil and the effect of thermal contraction, as the temperature is lowered. The static field dependence up to 3 Tesla of the inductive coil at T = 100K, is shown in the inset of Fig. 5. The overall frequency shift in this case is less than 50Hz indicative of a nearly field-independent flat response as expected for a non-magnetic metallic coil. Both the temperature and field dependence of the frequency shift for the coil are quite repeatable and not very different for the two cases viz. coil axis parallel or perpendicular to the DC field provided by the PPMS. As a precaution, it should be noted that if the coil is not rigidly mounted, one may have to contend with movement of the coil due to Lorentz force acting on it when the oscillating RF field (and hence the RF current) is perpendicular to the applied static field.

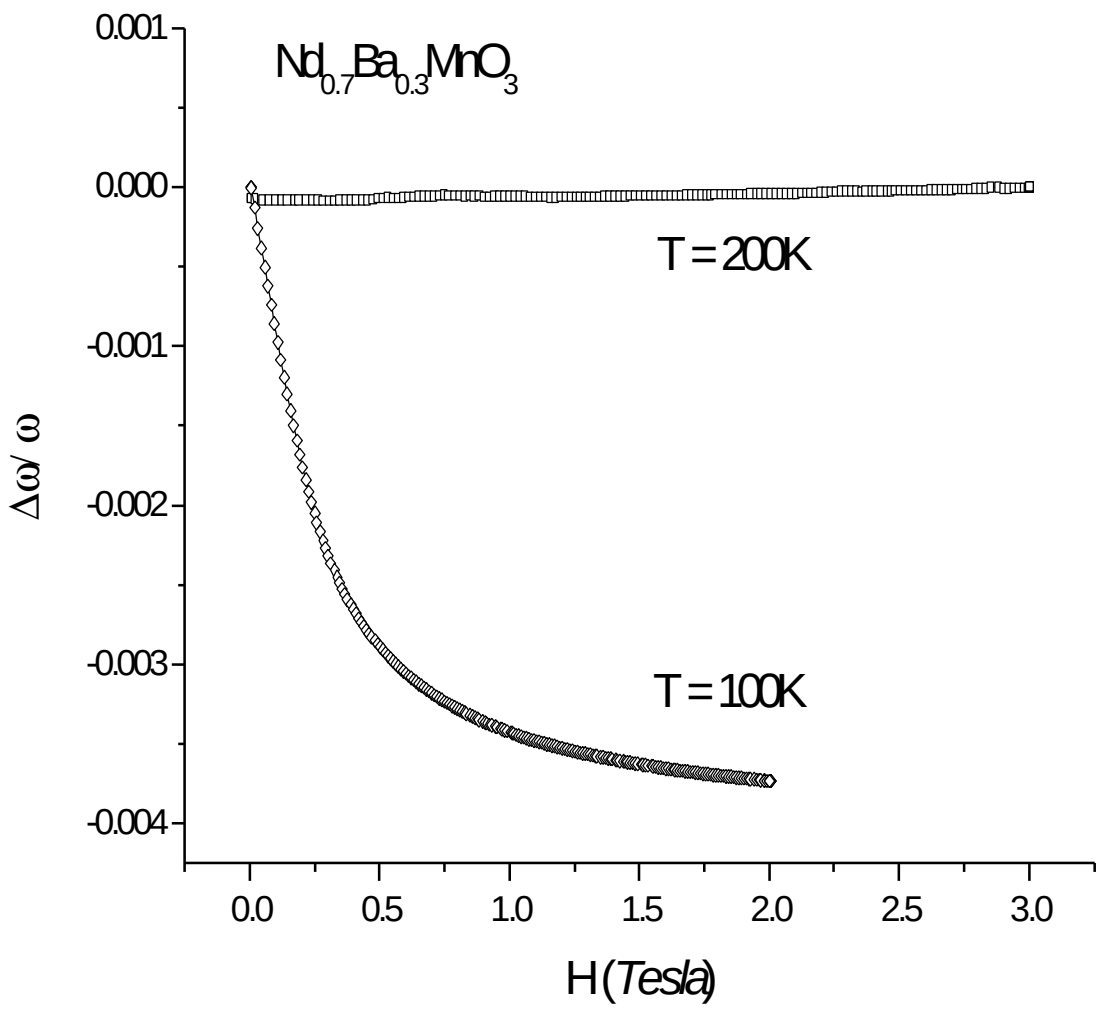

Fig. 7 Field dependence of RF shift for the Nd-Ba-Mn-O sample above and below the ferromagnetic $T_c$. Note the large negative magneto-impedance for T = 100K.

The TDO measurements on a manganese-based perovskite oxide sample, $Nd_{0.7}Ba_{0.3}MnO_3$, are presented in Figs. 6 and 7. This system was recently synthesized and its magnetic and magnetotransport characteristics were studied at our Institute [10]. The sample undergoes a paramagnetic to ferromagnetic transition as the temperature is lowered with the Curie temperature $T_c$ around 115K. In addition, the system exhibits colossal magneto-resistance (CMR) in the vicinity of $T_c$ [11].

In the TDO experiment, the $Nd_{0.7}Ba_{0.3}MnO_3$ material in powdered form is encapsulated in a plastic capsule that fits snugly into the core of the inductive coil L. Fig. 6 shows the temperature dependence of the RF shift in zero field. Note that the frequency shifts are much larger than the background shifts due to the empty coil and this has been eliminated in the data presented in Figs. 6 and 7. The paramagnetic to ferromagnetic transition is distinctly seen and the data is consistent with a large increase in the real part of the complex permeability (μ') at this transition. The field
dependence for the same sample is plotted in Fig. 7 where we have shown the data for two cases where the temperature is held above and below $T_c$. The striking difference in the RF impedance variation in the paramagnetic and ferromagnetic phases of the sample is quite obvious. For $T < T_c$, the RF impedance smoothly and monotonically decreases with increasing applied field and shows a tendency to saturate at higher fields. We ascribe this change to be directly associated with the magneto-impedance (MI) which is dominated by rapid change in the permeability. The shape of the curve and the saturation fields are different from that seen in the negative magneto-resistance associated with the CMR response. Detailed interpretation of the results and a comparison of the MI and MR effects for this system are beyond the scope of this instrumentation paper and will be discussed in a forthcoming publication [12].

In summary, we have designed and set up a very sensitive resonant experiment based on a TDO method, to study RF impedance changes in materials. The TDO circuit was tested at various stages and the system has been integrated with a customized low temperature cryostat to operate with a commercially available PPMS. Preliminary results on CMR oxides indicate that this instrument provides a novel way to study the spin and charge dynamics in materials.

## 6. ACKNOWLEDGEMENTS

This work at AMRI was supported through DARPA Grant No. MDA 972-97-1 0003. The authors would like to thank Drs. K-Y. Wang and J. Tang for providing the $Nd_{0.7}Ba_{0.3}MnO_3$ sample and Dr. C. J. O'Connor (Director of AMRI) for his strong encouragement and support for this research program. One of the authors (HS) is indebted to Dr. Sridhar of Northeastern University for first introducing him to the TDO technique and would like to acknowledge stimulating discussions with members of the Sridhar group.

---